# Ambipolar blends of CuPc and $C_{60}$: charge carrier mobility, electronic structure and its implications for solar cell applications

*Wolfgang Brütting\*, Markus Bronner, Marcel Götzenbrugger, Andreas Opitz*

Institute of Physics, University of Augsburg, 86135 Augsburg, Germany
E-mail: Wolfgang.Bruetting@physik.uni-augsburg.de

Summary: Ambipolar transport has been realised in blends of the molecular hole conductor Cu-phthalocyanine (CuPc) and the electron conducting fullerene $C_{60}$. Charge carrier mobilities and the occupied electronic levels have been analyzed as a function of the mixing ratio using field-effect transistor measurements and photoelectron spectroscopy. These results are discussed in the context of photovoltaic cells based on these materials.

## Introduction

In recent years organic semiconductors have attracted considerable interest due to their growing potential as active materials in electronic and optoelectronic devices. A long-standing paradigm, however, has been seen in their unipolar transport of electrical charges. Thus multi-layer structures comprising different organic materials with spatially separated electron and hole transport layers where used for fabricating efficient organic light-emitting diodes [1]. On the other hand in photovoltaic devices, owing to short exciton diffusion length in organic semiconductors, p- and n-conducting materials need to be in close contact which is usually realized by mixing them in one single layer yielding a so-called bulk-heterojunction structure [2-5]. Recently, such donor-acceptor mixtures have been implemented also in organic field-effect transistors (OFETs). Ambipolar OFETs have been realized with a variety of material combinations, including polymer/fullerene blends [6], mixtures of soluble oligomers [7] as well as evaporated molecular hetero-layer structures and mixed layers [8].

Our recent studies in this field have focussed on the combination of hole conducting copper-phthalocyanine (CuPc) with the electron conducting fullerene $C_{60}$. We have investigated OFETs with various mixing ratios and film preparation conditions and have demonstrated ambipolar inverters with these blends [9]. Using photoelectron spectroscopy, we have recently determined the occupied electronic levels and their changes upon mixing both materials [10]. In this contribution we will summarize these results and discuss their implications in the context of photovoltaic cells based on a combination of CuPc as electron donor and $C_{60}$ as electron acceptor, respectively. In particular we will demonstrate that differences in the open-circuit voltages of heterolayer cells and bulk-heterojunction devices can be traced back to the electronic structure of the blends.

**Film morphology and charge carrier mobility**

For characterisation in OFETs, films of neat CuPc and $C_{60}$ and their mixtures have been prepared by thermal evaporation on pre-structured Si wafers covered with thermally grown $SiO_2$. Details of the preparation can be found in Ref. [9]. Fig. 1 shows as an example the morphology of a 1:1 mixture measured by non-contact scanning force microscopy. The image clearly demonstrates that one is not dealing with a molecular mixture of both materials but rather with a nano-phase separated structure. Note that the height scale is more than twice the nominal film thickness of 25nm. Detailed studies [9] show that the tendency for demixing is strongest for such a 1:1 mixture and decreases towards more asymmetric mixing ratios.

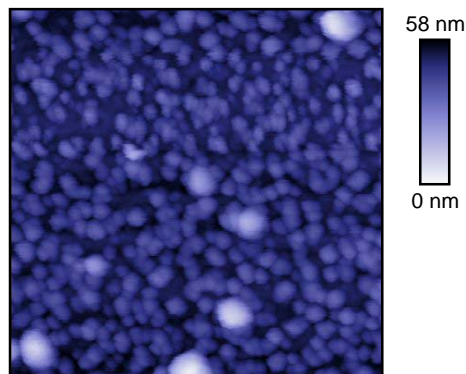

Figure 1: Scanning force microscopy image taken in non-contact mode for a 1:1 mixed film grown at a substrate temperature of 375 K. The total image size is $2\times2\mu m^2$. The film has an r.m.s. roughness of 6.5 nm.

We have also determined the field-effect mobility from the saturation regime of OFETs for both electrons and holes as a function of the mixing ratio. For these studies we have varied the substrate treatment (oxygen plasma or silanization of the $SiO_2$ surface) and the substrate temperature during film growth. The resulting mobilities are displayed in Fig. 2. Apparently, an exponential decrease of both $\mu_e$ and $\mu_h$ is observed upon dilution of the corresponding conducting material with the other species. A similar decrease has been reported in the literature for an ambipolar light-emitting system of co-evaporated molecular materials [11]. Furthermore, we find that heating the substrate during evaporation in combination with silanization increases the mobilities of neat films and mixed layers by up to four orders of magnitude. Remarkably, electron mobilities exceeding 1cm$^2$/Vs can be achieved in neat $C_{60}$ films. We also note that independent of the film growth conditions balanced electron and hole mobilities are achieved at about 25% fullerene content in the mixture, which is an important prerequisite for inverters with symmetric switching behaviour [9].

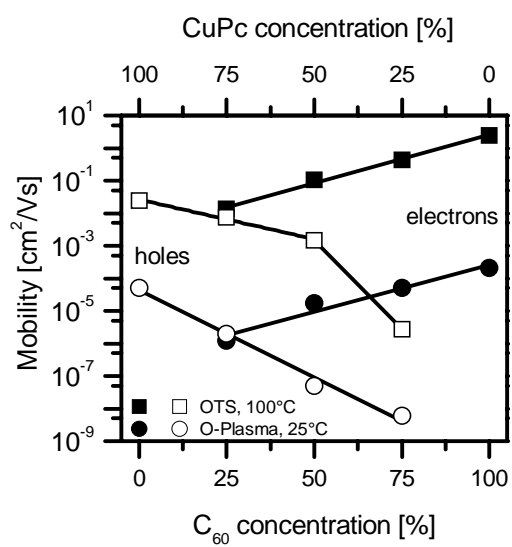

Figure 2: Electron and hole mobilities determined from the saturation regime of OFETs as function of the mixing ratio. The filled symbols are related to the electron transport, the open symbols to the hole transport.

**Electronic structure**

Using photoelectron spectroscopy the occupied electronic levels of neat films and mixtures of both materials have been determined with respect to the Fermi energy of an underlying gold substrate (for details see Ref. [10]). It was found that the energies of the highest molecular orbitals (HOMO), the core levels and the vacuum level vary linearly with the mixing ratio. Fig. 3 shows the synopsis of these studies, viz. the dependence of the HOMO and lowest unoccupied molecular orbital (LUMO) levels of both materials as a function of the composition of the film. Thereby one should note that the LUMO level has not been measured so far in blends, but we have taken the data from inverse photoemission for neat films of CuPc and $C_{60}$ and extrapolated towards blends assuming the same shift of the LUMO levels that was observed for all other levels [10]. In other words, we have assumed that the energy gaps of both materials stay constant in the mixture. This seems to be a reasonable assumption, as there is no indication for charge transfer between both materials in the ground state that would lead to new features in the electronic structure.

Postulating that this model is correct, one can draw some interesting conclusions from Fig. 3. First, it predicts that the injection barrier for holes into CuPc, i.e. the difference between the HOMO of CuPc and the Fermi level of Au (which is the reference level at binding energy zero in Fig. 3) is larger than the barrier for electron injection into $C_{60}$, which is actually observed in OFETs [9]. Further both injection barriers should decrease from neat films towards blends. However, if the contact resistance in OFETs is analyzed systematically (for details see Ref. [12]), one finds just the opposite behaviour: the contact resistance increases from neat films to blends. We have recently shown that this behaviour is indicative of diffusion limited injection where the decreasing mobility in blends (see Fig. 2) is responsible for an increase of the contact resistance [12].

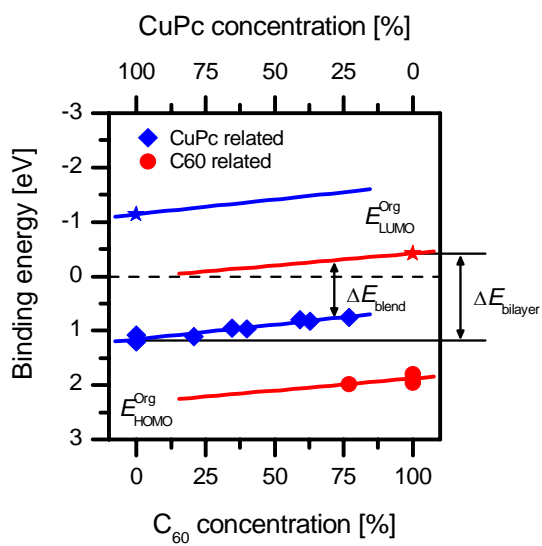

Figure 3. Position of the HOMO and LUMO levels as determined from photoelectron spectroscopy as a function of the $C_{60}$/CuPc mixing ratio. The solid lines are linear fits of the measured values, the constant dashed line is the Fermi level of the Au substrate. The LUMO levels are calculated from the transport gap of the neat materials assuming a constant electron affinity.

## Photovoltaic cells

Another consequence of the electronic level scheme shown in Fig. 3 is related to photovoltaic cells based on CuPc and $C_{60}$ as a donor-acceptor system. Upon photoexcitation excitons generated in either of the two materials (within the reach of the respective exciton diffusion length) will be dissociated at the organic-organic interface leading to electrons in the LUMO of $C_{60}$ and holes in the HOMO of CuPc. This charge separation leads to a gradient of the chemical potential at the interface that drives the photocurrent through the cell [13]. Thus the magnitude of the HOMO-LUMO offset between both materials will be related to the driving force, for which the open-circuit voltage can serve as an easily accessible experimental quantity. As already indicated by the arrows in Fig. 3, the obtained electronic structure therefore allows the conclusion that the HOMO-LUMO offset in blends is significantly smaller than at the heterojunction between neat layers of CuPc and $C_{60}$.

To test this prediction we have measured the photovoltaic response of photodiodes with indium-tin oxide (ITO) anodes covered with 30nm of the conducting polymer PEDOT:PSS (BAYTRON P), an 80nm thick active organic layer and a LiF:Al cathode. Thereby we have compared a bilayer structure comprising 40nm of CuPc and 40nm of $C_{60}$ on the one hand and an 80nm thick 1:1 mixture of both materials on the other hand. The obtained current-voltage characteristics in the fourth quadrant are shown in Fig. 4 for different light intensities. It can be seen that the blend system has significantly higher short-circuit currents, but the built-in voltage where all curves cross each other is smaller by 0.25V. For clarification we have plotted the open-circuit voltage in dependence of the light intensity in Fig. 5. Evidently, the open-circuit voltage in the blend is by 0.15V smaller and the difference is more or less independent of the light intensity. We note that the observed behaviour is in excellent agreement with recent numerical simulations by Marsh et al. [14].

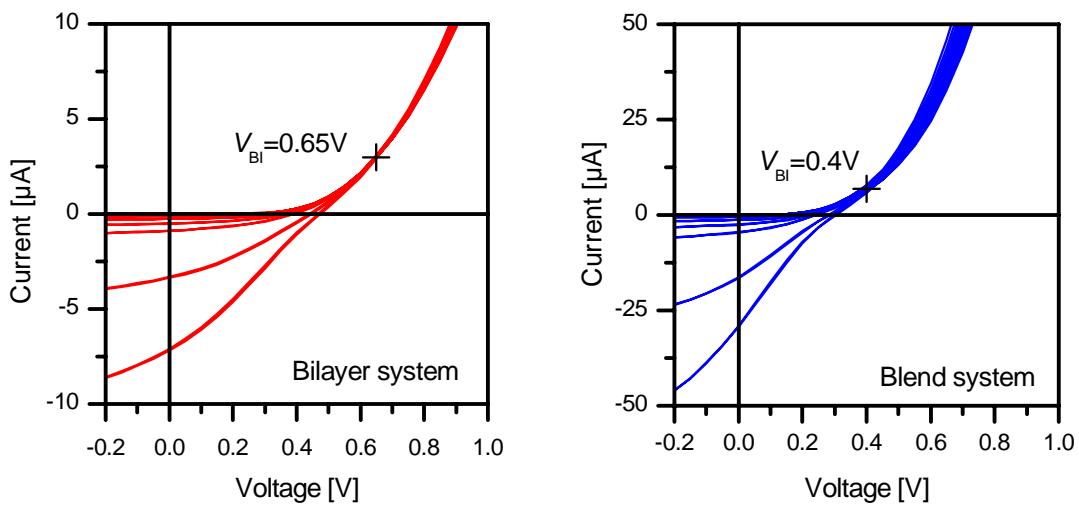

Figure 4. Current-voltage characteristics for a bilayer (left) and a blend (right) solar cells with the device structure ITO/PEDOT:PSS/organic layer(s)/LiF:Al. The curves are shown

for different light intensities. The built-in voltage is marked as the crossing point of all curves.

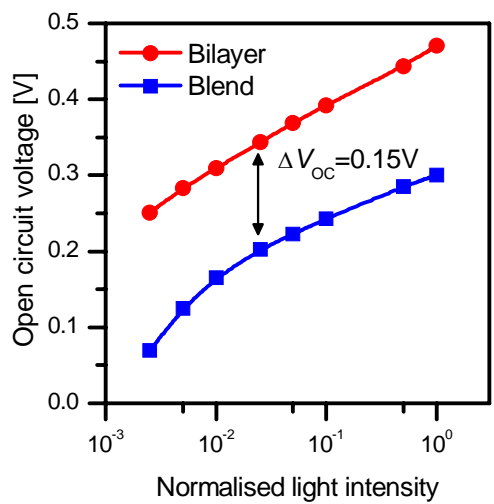

Figure 5. The open circuit voltage of the cells shown in Fig. 4 as a function of the light intensity (the maximum light intensity was about 20mW/cm$^2$).

## Conclusion

We have investigated charge carrier mobility and electronic structure of neat films and mixtures of CuPc and $C_{60}$ and have discussed the implications for photovoltaic cells based on these materials. We demonstrate that the bulk-heterojunction concept, though being very successful by providing a large active volume for exciton dissociation, has two weaknesses: (1) a reduction of charge carrier mobility and (2) a lower open-circuit voltage as compared to heterolayer structures. It will be a challenging task for future studies to control morphology and nano-phase separation on a length scale suitable for both exciton diffusion and charge carrier transport to further improve molecular solar cells.

---

Financial support by the Deutsche Forschungsgemeinschaft through Priority Programme 1121 is gratefully acknowledged.